\documentclass[10pt]{article}

\usepackage{amsmath}
\usepackage{amssymb}

\usepackage{graphicx} 

\usepackage{cite}

\usepackage{color} 

\usepackage[labelfont=bf,labelsep=period,justification=raggedright]{caption}

\makeatletter
\renewcommand{\@biblabel}[1]{\quad#1.}
\makeatother

\newcommand{\Knet}{$K^{net}$}
\newcommand{\Knode}{$K^{node}$}

\date{}

\usepackage{todonotes}
 \usepackage{newfloat}
\DeclareFloatingEnvironment[name={Supplementary Figure}]{suppfig}
\DeclareFloatingEnvironment[name={Supplementary Table}]{supptable}

\begin{document}

\begin{flushleft}
{\Large
\textbf{SANTA: quantifying the functional content of molecular networks}
}
\\
\medskip
Alex J. Cornish$^{1}$ and 
Florian Markowetz$^{2\ast}$
\\
\medskip
{\small
{\bf 1} Department of Life Sciences, Imperial College London, London, United Kingdom
\\
{\bf 2} Cancer Research UK Cambridge Institute, University of Cambridge, Cambridge, Cambridgeshire, United Kingdom
\\
$\ast$ E-mail: florian.markowetz@cruk.cam.ac.uk
}
\end{flushleft}

\section*{Abstract}
Linking networks of molecular interactions to cellular functions and phenotypes is a key goal in systems biology. 
Here, we adapt concepts of spatial statistics to assess the functional content of molecular networks.
Based on the guilt-by-association principle, our approach (called SANTA) quantifies the strength of association between a gene set and a network, and functionally annotates molecular networks like other enrichment methods annotate lists of genes.
As a general association measure, SANTA can (i)~functionally annotate experimentally derived networks using a collection of curated gene sets, and (ii)~annotate experimentally derived gene sets using a collection of curated networks, as well as (iii)~prioritize genes for follow-up analyses.
We exemplify the efficacy of SANTA in several case studies using the \emph{S. cerevisiae} genetic interaction network and genome-wide RNAi screens in cancer cell lines.
Our theory, simulations and applications show that SANTA provides a principled statistical way to quantify the association between molecular networks and cellular functions and phenotypes.
SANTA is available from http://bioconductor.org/packages/release/bioc/html/SANTA.html.

\section*{Author Summary}
Molecular networks are maps of the tens of thousands of interactions that occur between the components of biological systems. 
Types of interactions include physical, genetic and functional interactions between genes, gene products and metabolites. 
Network-based approaches to molecular biology are increasingly being used to better understand cellular functions.
Currently, gene set methods can be used to functionally annotate the hits from high-throughput studies; however, no methods exist to functionally annotate molecular interaction networks. 
This greatly limits our ability to quantify the often subtle functional adaptions that occur in networks as they re-wire to respond to external stimuli. 
Here, we extend well-tested concepts from spatial statistics to define a general association measure between networks and gene sets. 
Like Gene Set Enrichment Analysis, our approach measures concordant changes, but does this on networks, rather than on lists of genes. 
We validate it both in simulations and real-world case studies.
We apply our approach to genetic interaction networks mapped under different conditions and created using different methods, and demonstrate how it extends the previous analyses of data sets, allowing us to better understand the high-level changes that occur within cells.

\section*{Introduction}
High-throughput studies, like measuring differential expression in RNA-seq experiments or morphological changes in RNAi screens, produce genome-wide data that are often difficult to interpret. 
Functional annotation of hits in such studies relies mostly upon gene set analysis methods, which measure an association between hits and pre-defined gene sets~\cite{Ashburner2000,Subramanian2005} 
by quantifying overlap \cite{Beissbarth2004} or concurrent trends~\cite{Subramanian2005}.
These approaches generally treat hits as independent; only very few exploit an internal structure~\cite{Khatri2012}.
Recently, many high-throughput studies have produced networks of physical~\cite{Vidal2011,Rozenblatt-Rosen2012} or genetic~\cite{Costanzo2010,Ryan2012,Laufer2013} interactions, rather than lists of hits. 
These networks can be even harder to interpret than lists of hits. 
While more and more networks are being generated, rigorous statistical methods to annotate their functional content are lacking, thereby making it difficult to identify and quantify any high-level changes. 
The need for rigorous functional analysis of networks becomes especially evident when trying to quantify the often subtle functional adaptions observed in networks specific to external stimulation~\cite{Bandyopadhyay2010,Guenole2013}, different phenotypes~\cite{Laufer2013}, cell types~\cite{Kiel2013,Wang2006}, or diseases~\cite{Pujana2007,Sakai2011}.



\begin{figure}[!ht]
\begin{center}
\includegraphics[width=14cm, trim= 0mm 0mm 0mm 0mm]{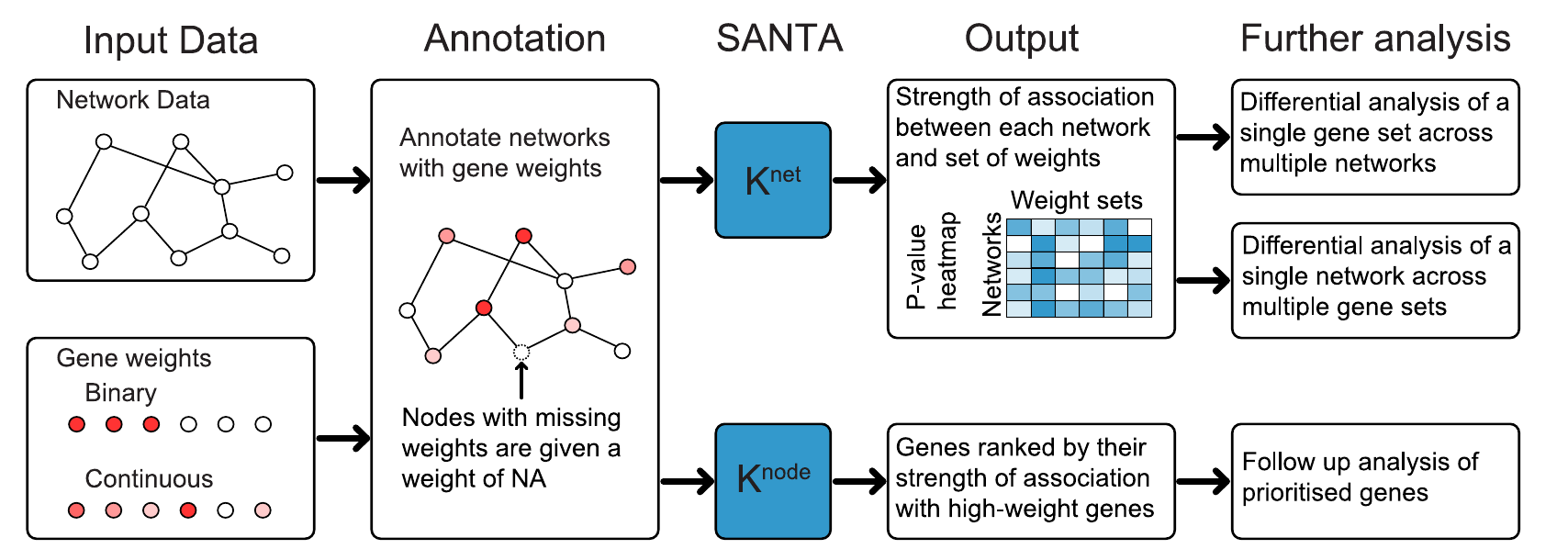}
\end{center}
\caption{
{\bf Overview of SANTA.}  
SANTA can be used both to quantify the strength of association between networks and sets of node weights (using \Knet) and to prioritise gene for follow-up analyses (using \Knode). Different node colour intensities represent different node weights. 
}
\label{fig:pipeline}
\end{figure}

Here, we develop methodology, called SANTA, for the rigorous and unbiased functional annotation of molecular networks.
The basic input to SANTA are a network and a gene set and the output is the statistical significance of their association (Figure~\ref{fig:pipeline}).
Like Gene Set Enrichment Analysis~\cite{Subramanian2005}, SANTA measures concordant changes in phenotype, but extends this concept to networks rather than lists of genes.
Our work is directly motivated by a study describing the functional content of the \emph{S. cerevisiae} genetic interaction network by Costanzo~\emph{et al.}~\cite{Costanzo2010}.
The iconic first figure of this paper  shows a network connecting genes with similar genetic interaction profiles and nodes highlighted according to their membership to functional groups defined by the Gene Ontology.
Costanzo~\emph{et al.} find that genes displaying tightly correlated profiles form discernible clusters corresponding to distinct bioprocesses and that the relative distance between distinct clusters appears to reflect shared functionality \cite{Costanzo2010}.
This is an important observation, because it shows which cellular functions are associated with the genetic interaction network.
If the network had been generated under different experimental conditions that activate different processes in the cell, these functional associations would most probably have changed.
Using SANTA, it is possible to quantify the significance of these changes in functional association in a principled statistical way, something not previously possible.

\begin{figure}[!ht]
\begin{center}
\includegraphics[width=14cm, trim= 0mm 0mm 0mm 0mm]{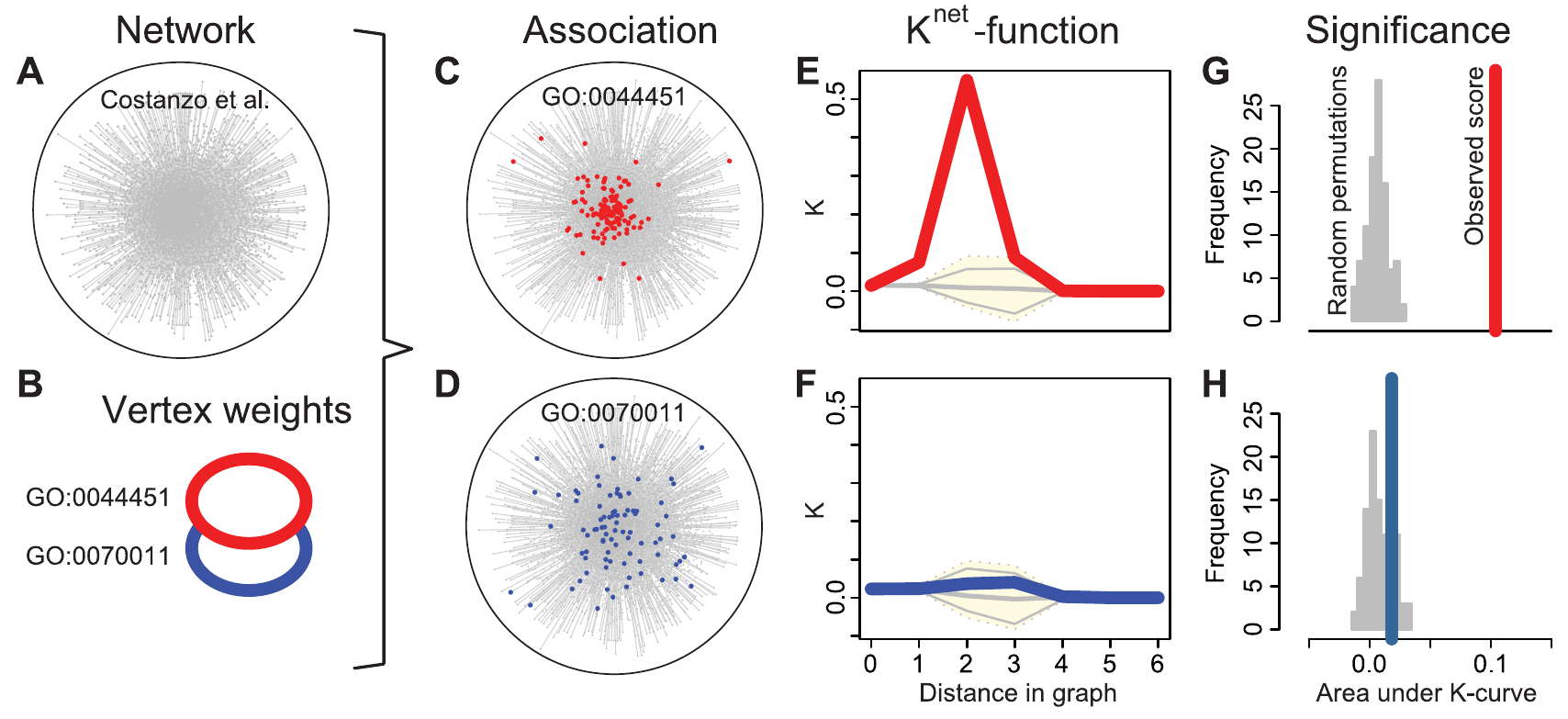}
\end{center}
\caption{
{\bf Application of the \Knet-function to two gene sets.} 
Example input: (A) {\it S. cerevisiae} GI map and (B) gene sets obtained from GO (`GO:0044451: nucleoplasm part' and `GO:0070011: peptidase activity'). 
(C, D) Network annotated with each gene set. 
From visual inspection, it appears that the gene set in (C) clusters more significantly than the gene set in (D). 
SANTA allows us to assess this clustering objectively. 
(E, F) The \Knet-function is computed for the observed gene sets (red and blue lines) and for a large number of permutations (yellow area). 
(G, H) In order to quantify the significance of the clustering, the area under the \Knet-function curve (AUK) is computed for the observed gene set (red and blue lines) and for each permutation (grey histogram). 
An empirical p-value is calculated using a Z-test. For GO:0044451, $p = 5.680 \times 10^{-30}$ and for GO:0070011, $p = 0.174$, demonstrating objectively that the gene set in (C) does cluster more significantly.
}
\label{fig:go_term_example}
\end{figure}

\paragraph{SANTA rigorously implements an intuitive association measure}
The roadmap for functional analysis of networks provided by Costanzo~\emph{et al.}~\cite{Costanzo2010} relies on assessing the clustering of selected nodes on the network.
However, their analysis was done by eye and depends not only on the gene set and the network but also on the visualisation algorithm used.
Clustering on a network is an intuitive measure of functional content, but without rigorous statistical methods the significance of observed patterns can not be assessed objectively.
To address this problem, we have adapted well-tested concepts from spatial statistics \cite{Ripley1981} to define an objective and quantitative association measure between networks and gene sets.
SANTA (\underline{s}patial \underline{a}nalysis of \underline{n}e\underline{t}work \underline{a}ssociations) is built on the guilt-by-association principle:  if a gene set shows a surprising degree of clustering on a network, we call them `associated' (Figure~\ref{fig:go_term_example}C); if the gene set is randomly distributed over the network, we call them `not associated' (Figure~\ref{fig:go_term_example}D). 

\paragraph{Previous research}
Enrichment analysis is a well-developed field and Khatri~\emph{et~al}~\cite{Khatri2012} recently described three `generations' of statistical methods, from over-representation analysis and functional class scoring, to pathway-topology based approaches.
While methods in the last category \cite{Rahnenfeuhrer2004,Draghici2007,Shojaie2010} use pathway topology to compute gene-level statistics, none of them can be directly applied to measure the functional content of a network. 
Two related approaches, the compactness score of PathExpand \cite{Glaab2010} and EnrichNet \cite{Glaab2012}, compare the clustering of two sets of genes on a network, but not the significance of the clustering of a single set. 
While the compactness score can be adapted to measure the significance of clustering, it focusses on the local structure of the network and can be less effective than SANTA to detect global effects, as we show below. 
Other approaches overlay interaction networks with genome-wide measurements and identify enriched subnetworks \cite{Ideker2002,Sanguinetti2008,Beisser2010}, to which enrichment analysis can then be applied in a consecutive step \cite{Jia2011}.
Again, subnetwork identification does not directly measure the association between gene sets and networks, and we show the effect of this difference in a comparison study.

In summary, no approach currently exists that, like SANTA, globally assesses the functional content of a network. 
In the following, we describe the methodology underlying SANTA and test its efficacy by applying it to both simulated and real data. 
Gene set enrichment methods have had a big impact on biological research and are used in almost every analysis of experimentally derived gene lists. 
The case studies we present in this paper show that SANTA has the potential to have a similar impact on all functional studies of network data.

\section*{Results}

\subsection*{Adapting Ripley's K-Function for networks}
Spatial statistics model spatial correlation structures between observations (analogous to how time series analysis models the correlation between time points)~\cite{Gaetan2010}.
One area of spatial statistics analyses point patterns and asks if points in $R^2$ are occurring at random or cluster together in any way. 
A  basic tool for the analysis of point patterns is Ripley's $K$-function~\cite{Ripley1981}, which can be estimated by computing how many other points are captured by drawing a circle of radius $s$ around each point:

\begin{equation}
K(s) \ = \  \frac{1}{\lambda n} \sum_i \sum_{j \not= i} \mathbf{I}(d(i,j) < s)
\label{K}
\end{equation}

where $n$ is  the number of points, $\lambda$ is the density (number of points per unit area),  $d(i,j)$ is the distance between two points in $R^2$, and $\mathbf{I}( d(i,j) < s)$ is an indicator function with value $1$ if the distance $d(i,j)$ between points $i$ and $j$ is smaller than $s$, and $0$ otherwise.
If the points are densely clustered, most of them will be found for small values of $s$, while for uniformly spread points the $K(s)$ function achieves larger values only for large values of $s$.
Applications of the $K$-function in computational biology include detecting host factors involved in virus infection by observing the clustering of infected cells in siRNA screening images \cite{Suratanee2010}.

In our scenario, we measure observations at fixed locations (the nodes in the graph) instead of random locations in the plane. 
However, we can still adapt basic spatial statistics methodology:
In the following paragraphs we will define a $K$-function for weighted networks, called \Knet, by first defining distance measures on graphs (instead of $R^2$) and then weighting each node by the strength of its phenotype.
To adapt $K(s)$ to networks, we formalised the problem using a node-labeled and edge-weighted graph $G = (V,E)$, where $V$ is a set of $n$ nodes (vertices) and $E \subseteq V \times V$ a set of $m$ (undirected) edges between pairs of nodes $(i,j)$. 
Node weights $\{ p_v \ | \ v \in V \}$ correspond to the strength of a gene's phenotype ($p_v \in R$,  $v \in V$) or whether the gene is part of a particular functional group ($p_v \in \{0,1\}$,  $v \in V$) and edge weights $\{ w_e \ | \ e \in E \}$ correspond to the strength or significance of interactions. 

\paragraph{Graph distances}
Distances between non-neighbouring nodes in this graph can be measured in many ways, including shortest path lengths, diffusion kernels \cite{Kondor2002} and the mean first-passage time \cite{White2003}, which are all implemented in the SANTA software package.
There are subtle differences in the aspects of the network structure incorporated within each measure. 
For example, a shortest-path approach will only take into consideration the one path with the shortest length, no matter how many other paths exist between two nodes.
A diffusion kernel, on the other hand, takes into account all paths and will yield a smaller distance the better connected two nodes are.
The results produced by SANTA are generally robust across distance measures (Figure S\ref{suppfig:distance_robustness}), meaning that it often does not matter which method is chosen by the user. 
The shortest path distance method requires the least computational time and therefore we will mainly use this method in the paper. 
Efficient algorithms like Dijkstra's or Jonsohn's exist to compute shortest paths between all pairs of nodes \cite{Cormen2009} and are conveniently implemented in software packages like `igraph' \cite{Csardi2006}.
However, the diffusion-kernel based distance measure is used to identify enriched subnetworks, as this method is seen to produce denser subnetworks. 

Many of the graph distance algorithms assume that small edge-weights correspond to stronger functional association between the two nodes. 
Many networks, however, are built by correlation analysis, where stronger functional associations are shown by a larger weight.
Thus, in practice, the edge weights in a given molecular network often need to be reweighed to be used as graph distances $\{ d_e \ | \ e \in E \}$. 
Due to differences between the methods used to create each molecular network, it is necessary to use a different approach when reweighing the edges of each network. 
Edges are reweighed so that the strongest interactions have a graph distance of 0 and the weaker the interaction the greater the distance (see Methods).

\paragraph{Node weights}
Exchanging the planar distance $d(\cdot,\cdot)$ in Equation~(\ref{K}) with a graph distance $d^{g}(\cdot,\cdot)$ directly results in a version of the $K$-function that is applicable if the node weights are in $\{0,1\}$.
However, in many real situations, e.g. differential expression analysis or large-scale RNAi screens, the node weights are real numbers.
In this case it is not only of interest \emph{how many} `hits' are close to each node, but also \emph{how strong} these hits are.
We implement this notion by weighting the contribution of each node by the relative weight it carries compared to the other nodes.
This results in a function \Knet{} of the form

\begin{equation}
\label{Knet}
K^{net}(s) = \frac{2}{(\bar{p} n)^{2}}\sum_{i}p_{i}\sum_{j}(p_{j} - \bar{p})~\textbf{I}(d^{g}(i,j)\leq s) 
\end{equation}

where $p_i$ is the phenotype observed at node $i$ and $\bar{p} = \frac{1}{n}\sum_{i=1}^n p_i$.
This re-weighting is very similar in spirit to the re-weighting of the Kolmogorov-Smirnov statistic in GSEA \cite{Subramanian2005}.
Generally, we plot \Knet{} from $0$ to the maximal distance within the graph (the diameter), in which case \Knet{} forms a curve starting and ending at $0$ (Figure S\ref{suppfig:ripley_vs_knet}).

\paragraph{Node-wise K-function}
The inner sum of Equation~(\ref{Knet}) offers a natural way to prioritise nodes and identify candidate genes for the mechanisms underlying the observed phenotype. 
We define this node-wise $K$-score as the AUK of the node-wise $K$-function, defined as:

\begin{equation}
\label{Knode}
K_{i}^{node}(s) = \frac{2}{\bar{p} n}\sum_{j}(p_{j} - \bar{p})\ \textbf{I}(d^{g}(i,j)\leq s) 
\end{equation}

\paragraph{Computing significance by permutation}
To check how significant observed \Knet{} results are, we compare them to curves obtained by applying \Knet{} to sets of randomly permuted hits. 
These sets of permuted hits are obtained by randomly redistributing the node weights across the nodes. 
When permuting the node weights, it is not always possible to maintain the degree of each node, therefore, node degree is not considered when permuting the weights.

Since we want to quantify the amount of clustering, we are interested in observed \Knet-curves that ascend steeper than random curves.
To quantify this, we compute for all curves (the observed \Knet{} and the $N_{perm}$ random permutations) the \emph{area under the \Knet -curve}, or \emph{AUK} value.
An empirical p-value for the observed AUK is calculated using a Z-test. Figure~\ref{fig:go_term_example} exemplifies the application of \Knet{} to two GO terms and the yeast genetic interaction networks.

\subsection*{Simulation studies}

\paragraph{SANTA successfully identifies clustering on simulated networks}
Functional annotation is an exploratory task without a general gold standard. 
In order to test the ability of SANTA to correctly identify clustered distributions of node weights on networks, we conducted a number of controlled simulations. 
In each of these simulations, we created a network containing a cluster of high weight vertices of a known strength and applied the \Knet-function in order to determine whether it would successfully identify the clustering. 

Each of the networks contained 500 nodes and was created using the Barabasi-Albert model of preferential attachment \cite{Barabasi1999}. 
A seed node was chosen at random. 
All nodes in the network were ranked by their distance (using the shortest paths method) to the seed node and the $s$ closest nodes chosen to be the sample set. 
A hit set was the created by choosing 5 nodes at random from the sample set. 
Different values of $s$ (10, 20, 50, 100 and 500) were chosen to simulate different clustering strengths.
A value of $s$ equal to the number of nodes in the network is the same as randomly sampling nodes from the entire network. 

\begin{figure}[!ht]
\begin{center}
\includegraphics[width=8cm, trim= 0mm 0mm 0mm 0mm]{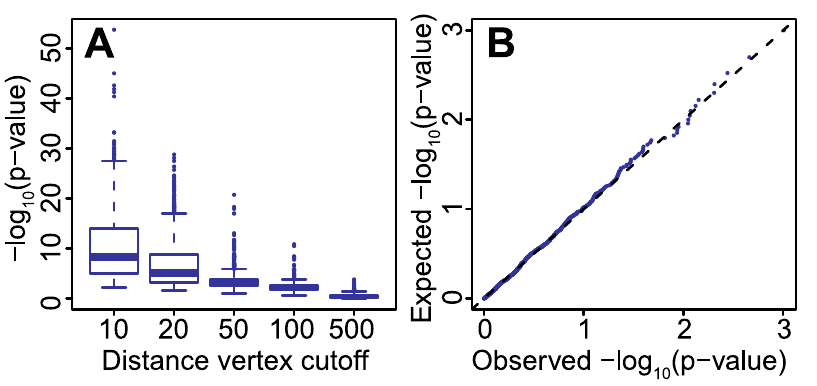}
\end{center}
\caption{
{\bf Application of \Knet~to simulated networks.}  
Scale-free networks containing clusters of high-weight nodes of various strengths were generated. 
The smaller the distance cutoff used to generate the cluster, the greater the strength of the clustering. 
(A) 1000 trials were completed for each distance cutoff. 
As expected, the most significant clustering was measured by the \Knet-function when smaller distance cutoffs were used. 
(B) Q-Q plot of the p-values observed in the simulation study trials in which no distance cutoff was used and the p-values expected under the uniform distribution. 
The high-weight nodes were distributed homogeneously when no distance cutoff was used. 
The observed p-values deviate little from the expected p-values, demonstrating that the \Knet-function does not detect clustering when clustering is not present.
}
\label{fig:simulation}
\end{figure}

As expected, SANTA identified more significant clustering when applied to hit sets created with smaller values of $s$ (Figure~\ref{fig:simulation}A). 
When nodes are randomly sampled from the entire network, the p-values returned by SANTA were uniformly distributed (Figure~\ref{fig:simulation}B), as expected when the null hypothesis is true.

\paragraph{SANTA incorporates the global structure of a network for functional association}
One of the main advantages of the \Knet-function is that it considers the global topology of a network when measuring the significance of clustering. 
This can be demonstrated by comparing the \Knet-function to an adapted version of the compactness score \cite{Glaab2010}.
The compactness score of a gene set is the mean distance between pairs of nodes in the gene set. 
It is used by the PathExpand tool to compare the clustering strength of different sets of nodes \cite{Glaab2010}. 
By comparing the compactness score of an observed set of nodes to the compactness scores of permuted sets of nodes, it is possible to produce an empirical p-value describing clustering significance, much like the \Knet-function.

\begin{figure}[!ht]
\begin{center}
\includegraphics[width=8cm, trim= 0mm 0mm 0mm 0mm]{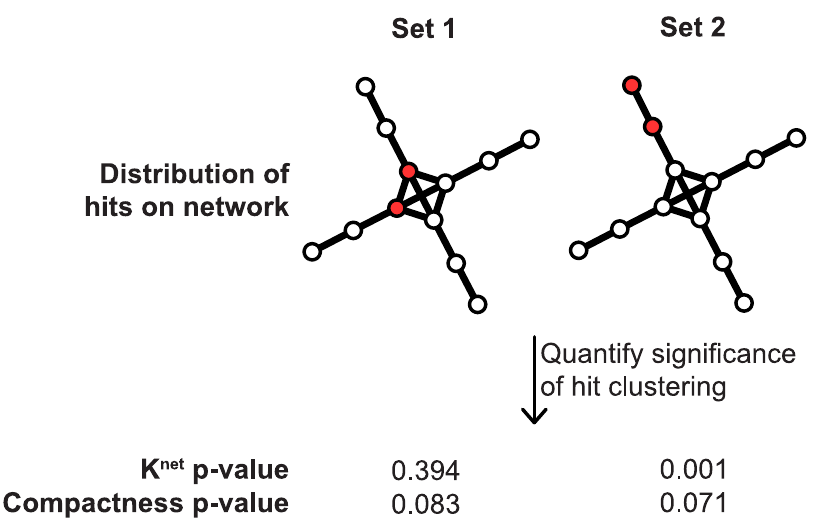}
\end{center}
\caption{
{\bf Comparison of \Knet~and Compactness.}
Example of the difference between the \Knet~and the Compactness functions.
Red circles represent hits on the network.
P-values were computed for both functions using 1000 permutations.
Only the \Knet-function incorporates the global structure of the network and therefore only it identifies a more significant association between set 2 and the network.
}
\label{fig:knet_vs_compactness}
\end{figure}

Many real-world networks follow a power-law degree distribution and contain nodes with both a small and large number of interacting partners \cite{Barabasi1999}. 
If the genes in a gene set all have a large number of interacting partners, then the presence of interactions between the genes in the gene set could be considered less significant, as there is a greater likelihood that they would be observed by chance. 
Therefore, it is necessary to incorporate the global structure of the network and consider the number of nodes located near each node when quantifying clustering significance. 
Figure~\ref{fig:knet_vs_compactness} demonstrates that while the \Knet-function incorporates the global structure of the network, the compactness score does not. 
The \Knet-function can also be applied to continuous distributions of node weights, while the compactness score can only be applied to binary sets, limiting its application. 
For these reasons, the \Knet-function is better suited to measuring the significance of clustering of node weights on real-world network. 

\paragraph{SANTA provides a complementary method of identifying enriched subnetworks}
Next, we compared SANTA to approaches that overlay molecular networks with additional node information and identify a high-scoring subnetwork, using simulated and real data. 
A widely used example is BioNet~\cite{Beisser2010},  which identifies enriched subnetworks of nodes by fitting a beta-uniform mixture (BUM) model to the network in order to score nodes. 
Positive-scoring nodes are then aggregated and a minimum spanning tree calculated between these positive nodes. 
However, the presence of negative-scoring nodes between clusters can prevent BioNet from identifying multiple clusters.
As the \Knode-function considers each node individually, it is able to return high-scoring nodes spread across multiple clusters.

\begin{figure}[!ht]
\begin{center}
\includegraphics[width=8cm, trim= 0mm 0mm 0mm 0mm]{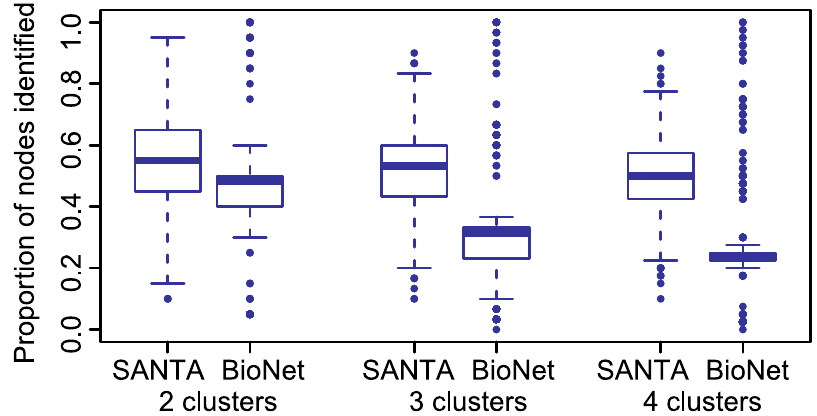}
\end{center}
\caption{
{\bf Comparison of \Knode~with BioNet.}
Comparison of the ability of the \Knode-function and BioNet to identify high-weight nodes contained within multiple clusters on a single simulated network. 
Across 1000 trials, the \Knode-function identified a greater proportion of the high-weight nodes when they were distributed across 3 or 4  clusters.
}
\label{fig:bionet_simulated}
\end{figure}

We conduced a number of simulations in order to compare the abilities of \Knode{} and BioNet to identify high-scoring nodes located within multiple clusters on a network.
In each simulation, a network containing 1000 nodes was created using the Barabasi-Albert model of preferential attachment \cite{Barabasi1999}. 
2, 3 or 4 nodes from distant parts of the network were selected to seed the clusters.
For each seed node, 10 nodes were selected at random under a probability distribution that ensured that the probability of being chosen decreased exponentially with the distance from the seed node ($P(i) \sim 10^{-d^g(k, i)}$, where $k$ is the seed node).
The selected nodes became the high-weight nodes and were assigned node weights from a truncated normal distribution with a mean of 0 and a standard deviation of $1 \times 10^{-6}$, within the interval $[0, 1]$.
Unselected nodes were assigned node weights from the uniform distribution, again within the interval $[0, 1]$.
\Knode{} and BioNet were then applied to the network.
If $x$ high-weight nodes are applied to the network, \Knode{} is said to have successfully identified a high-weight node if it is ranked within the top $x$ nodes. 
BioNet successfully identifies a high-weight node if it is contained within the returned enriched subnetwork. 
Figure~\ref{fig:bionet_simulated} shows that \Knode{} was able to successfully identify a greater proportion of labelled nodes than BioNet when 3 or more clusters were added to the network.
BioNet tended to successfully identify nodes from a single cluster, but missed nodes contained within others.
This highlights an advantage of SANTA over methods identifying a \emph{single} top scoring subnetwork.

\subsection*{Real-world case studies}

\paragraph{SANTA identifies functionally-informative enriched subnetworks}
We also compared the \Knode-function to BioNet by rerunning the validation experiment conducted by BioNet. 
Gene expression data from two subtypes of diffuse large B-cell lymphomas (DLBCL) was combined with survival data \cite{Rosenwald2002}.
P-values were produced using Cox regression and these were converted into node weights which were used to annotate the HPRD interaction network \cite{KeshavaPrasad2009}.
BioNet and the \Knode-function were applied in order to identify enriched subnetworks.
BioNet returned a module containing 38 genes and 49 interactions. 
In order to make a fair comparison, the 38 genes ranked highest by \Knode{} were chosen to form the \Knode{} module.  
This module is denser than the BioNet network and contains 86 interactions.
Only 7 genes were identified by both BioNet and \Knode{}. 
The BioNet module is enriched with genes involved in the oncogenic NF$\kappa$B pathway \cite{Beisser2010}. 
Fisher's exact test was used to identify functional gene sets enriched within the modules \cite{Huang2009a}. 
While the \Knode{} module is not enriched with NF$\kappa$B pathway genes, the cancer-associated GO term `regulation of apoptosis' was identified as the most strongly-enriched gene set ($p < 1\times10^{-7}$). 

These results demonstrate that the \Knode{} function represents a complementary method of enriched subnetwork identification. 
However, the main purpose we envision for SANTA is to annotate the functional content of networks and the next case studies focus on this task.

\begin{figure}[!ht]
\begin{center}
\includegraphics[width=14cm, trim= 0mm 0mm 0mm 0mm]{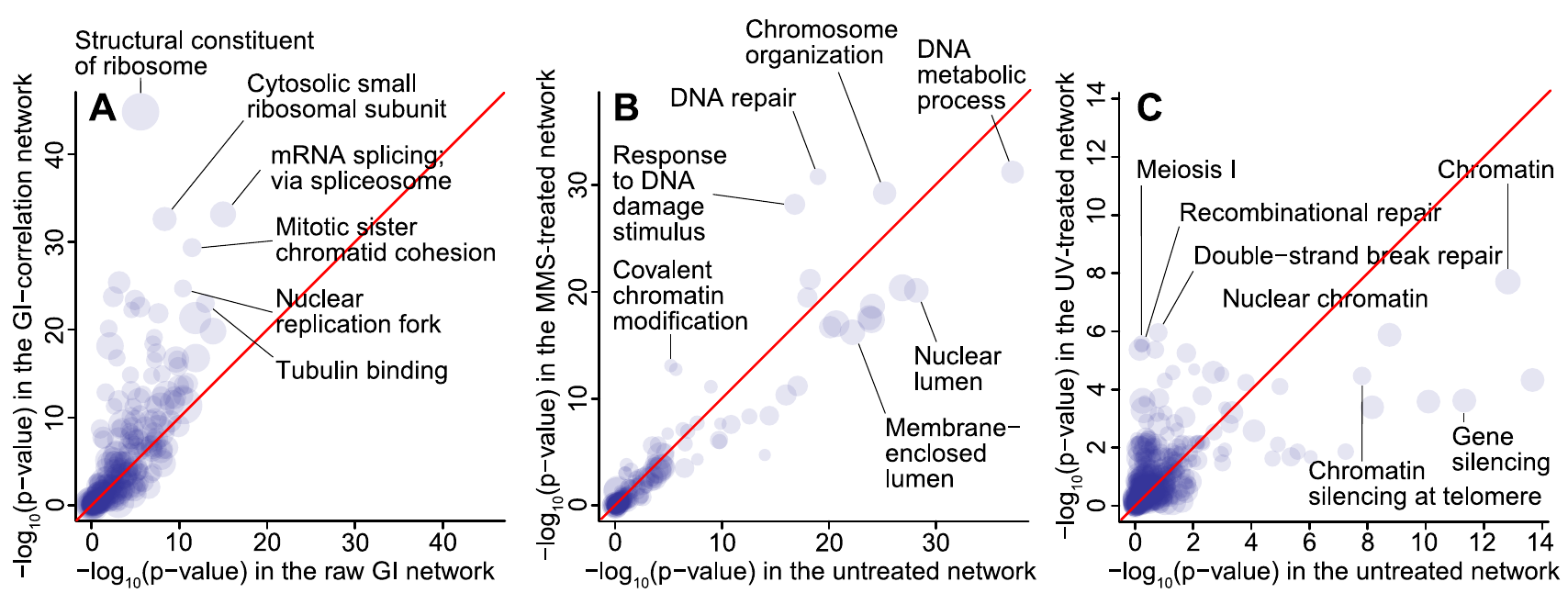}
\end{center}
\caption{
{\bf Applications of \Knet~to real networks.}  
 (A) Comparison of the functional content of a network of raw GIs and a network representing correlation in GI profile. 
 GO terms are associated more strongly with the GI-correlation network, indicating that this network is functionally more informative.
 (B) Comparison of the functional content of the untreated and MMS-treated GI networks. 
 GO terms associated with response to DNA damage were enriched within the treated network. 
 (C) Comparison of the functional content of the untreated and UV-treated GI networks. 
 GO terms associated with cell cycle progression were enriched within the treated network.
}
\label{fig:results1}
\end{figure}

\paragraph{Correlations in GI profile produce functionally more informative networks}
For further validation, we applied SANTA to the global genetic interaction (GI) network of \emph{S. cerevisiae}, where there is evidence that protein function is more closely related to the global similarity between GI profiles than to individual interactions \cite{Costanzo2010}.
To measure this effect we contrasted the functional content of a network of high correlations between GI profiles with a network of individuals GIs. 
This was done by quantifying the strength of association of sets of functionally related genes with each of the networks using the \Knet-function.
Sets of functionally related genes were obtained from the Gene Ontology (GO).  
To ensure that the functional sets were not too thinly or thickly spread, only GO terms associated with between 20 and 100 network genes were tested. 
Figure~\ref{fig:results1}A shows that GO terms indeed tend to cluster more strongly on the correlation network than on the network of individual GIs, demonstrating that similarity between GI profiles is a stronger indication of shared protein function. 
This effect was independent of the GO term size and strongest for specific cellular functions like `structural constituent of ribosome', `cytosolic small ribosomal subunit' and `piecemeal microautophagy of nucleus' (Table S\ref{supptable:raw_correlation}).

\paragraph{Yeast interaction networks functionally rewire under external stress}
Most studies have mapped GIs in cells under normal laboratory conditions \cite{Costanzo2010,Ryan2012,Frost2012}.
However, it has been demonstrated that GIs can be condition-dependant \cite{StOnge2007}. 
Mapping GI networks under multiple conditions is therefore likely to reveal new information about how a cell reorganises itself to cope with environmental conditions. 
To measure these effects, we used SANTA to analyse the changes in functional content that occur in \emph{S. cerevisiae} GI networks under external perturbation by the DNA-damaging agent methyl methane-sulfonate (MMS) \cite{Bandyopadhyay2010} and UV radiation \cite{Srivas2013a}.
We again used the association strength of GO term-associated gene sets to quantify functional enrichment within each network. 
By comparing the association strengths of the GO terms between the treated and untreated networks, it is possible to identify pathways and processes that are up- and down-regulated in response to the changes in environmental condition.
GO terms were applied to each network if they associated with between 20 and 100 genes. 

We found several GO terms that associated more strongly with the MMS-treated network than the untreated network (Figure~\ref{fig:results1}B and Table S\ref{supptable:dna_damage}).
GO terms related to the response to DNA-damage, including `DNA repair', `response to DNA damage stimulus' and `covalent chromatin modification', associated more strongly with the MMS-treated network. 
This result is expected and found in the original publication, thereby providing further validation for SANTA. 

Comparing the functional enrichment of the UV-treated network replicates the finding of the original publication as well as identifying subtle changes not reported in the publication (Figure~\ref{fig:results1}B and Table S\ref{supptable:uv}) \cite{Srivas2013a}.
The top 10 GO terms most strongly enriched within the UV-treated network are related to DNA-damage repair or cell cycle progression; processes known to be affected by exposure to UV radiation \cite{Sertic2012}. 
However, the \Knet-function is also able to identify processes affected by UV-treatment not reported in the original publication. 
`Chomatin silencing at telomere' associates more strongly with the untreated network ($p<1.6\times10^{-8}$) than the treated network ($p<3.4\times10^{-5}$).
It has previously been demonstrated that some of the proteins involved in transcriptional silencing at the telomeres, such as \emph{Sir} and \emph{Ku}, are also involved in DNA-damage repair \cite{Tham2002} and are dispersed from the telomeres in response to DNA damage \cite{Martin1999}. 
Our results provide further support for this hypothesis and demonstrate that the \Knet-function is able to provide insight into the functional repurposing of cells that cannot be provided by current methods. 

The strength of gene set association was independent of gene set size (Figure S\ref{suppfig:clustering_vs_size}).
Association strength is also robust across distance methods (Figure S\ref{suppfig:distance_robustness}).
SANTA identifies functional adaptions not seen in the original analysis and thereby also provides a method of hypothesis generation. 
The advantage of SANTA is that it directly contrasts the functional content of the two networks, which improves on the indirect enrichment analysis of differing edges in the original analysis \cite{Bandyopadhyay2010}.

\paragraph{Interaction networks provide different levels of information about cancer cell line maintenance}
Different networks describe different aspects of cellular machinery: co-expression networks describe transcriptional effects, protein interaction networks describe complexes and genetic interaction networks describe epistatic buffering relationships. 
Identifying the type of network that associates most with genes of interest can point to the mechanism underlying observed phenotypes.
To exemplify this idea, we used SANTA to associate RNAi screens in cancer cell lines \cite{Cheung2011} to a curated network of physical interactions \cite{Orchard2014} and to a functional interaction network created by combining 21 data sources from 4 species \cite{Lee2011}, with the aim of identifying the network that best explains the phenotype. 
The colon and ovarian cancer cell line RNAi hits were seen to associate more strongly with the functional interaction network (Figure~\ref{fig:results1}C), indicating that is is possible to create a network that better explains the mechanisms that maintain cancer cell line viability by combining multiple data source.

\section*{Discussion}
SANTA is a general approach for functional annotation that extends enrichment analysis from gene sets to networks.
SANTA combines the guilt-by-association principle, which is one of the most powerful paradigms for function prediction, with well-tested concepts adapted from spatial statistics. 
In this way, SANTA provides a rigorous implementation of an intuitive measure of functional annotation.
We have applied SANTA to several datasets from different organisms and our results show how SANTA rigorously addresses the basic question of which functional processes are reflected in a network. 

In yeast, our results on genetic interactions support the idea that a strong correlation of GI profiles between two genes is a greater indicator of shared function than the presence of a single GI. 
The reason for this increase in functional information is most probably that individual GIs don't bear much evidence for underlying mechanisms, while having many GI partners in common is strong evidence for genes acting in the same pathway or complex \cite{Tong2004}.
Additionally, Costanzo et al. \cite{Costanzo2010} noted that their network captured only 35\% of the previously reported interactions, indicating that a large number of false positives and false negatives may be present within GI networks. 
Networks of correlation in GI profile may be more robust to the high number of errors that are present when GIs are mapped. 

Extending these results to networks rewired under external stimulation, we show how SANTA quantifies subtle functional changes.
In human, we showed how SANTA can contribute to understanding the mechanisms underlying large RNAi screens. 
Testing the association of hits with many different networks (transcriptional, proteomic, genetic) can help us to understand which cellular mechanisms underly the phenotypes.
In summary, our results support that SANTA accurately quantifies the functional content of networks, points to mechanisms underlying observed phenotypes, and provides a natural way to compare functional changes across networks. 

We expect SANTA to contribute mostly to the functional annotation of networks derived under different environmental conditions (like the GI networks we used as case studies here).
However, SANTA is a very general approach and the examples we presented here also show other uses: it can also be used to annotate RNAi hits (if different functional networks are available) and prioritise individual hits over others (using \Knode).
In the future, we see many further opportunities for applying SANTA.
For example, new methods of automated, single-cell phenotyping measure genetic interaction networks across a broad spectrum of phenotypes~\cite{Laufer2013} and a functional annotation method like SANTA could have great impact on understanding which cellular processes are reflected in which phenotype.
Another potential application for SANTA lies in network-based medicine, where drug development for complex diseases is developing towards targeting dynamic network states \cite{Barabasi2011,Creixell2012,Erler2012} and network-based analysis has identified  cancer subtypes  \cite{Hofree2013}.
Functional annotation of these networks will further our understanding of the biology underlying these diseases.

Gene set enrichment analysis is the first step in the unbiased analysis of most experimentally derived gene lists and we expect SANTA to have a similar impact on all functional studies of network data.

\section*{Methods}

\subsection*{Shortest paths distance measure} 
There are a number of different methods available to calculate the distance between a pair of nodes in a network.
One of the simplest methods involves identifying the shortest path connecting the node pair and using the length of this path. 
The shortest paths distance measure can be applied to networks with or without weighted edges. 
In unweighted networks, the shortest path is equal to the number of edges included within the shortest path. 
In weighted networks, it is the sum of the edge weights along the shortest path. 

A number of different algorithms are available to compute the shortest path between two nodes. 
Which algorithm is most efficient depends on the type of network being analysed. 
If no edge weights are present, then the breadth-first search algorithm is ideal. 
If edge weights are present and each edge weight is non-negative, then Dijkstra's algorithm is more efficient \cite{Cormen2009}.

\subsection*{Diffusion kernel-based distance measure}
The diffusion kernel-based distance measure is another method of calculating distances between pairs of nodes \cite{Kondor2002}. 
An advantage of the diffusion kernel-based method over the shortest-paths method is that whilst the shortest-paths method calculates the distance along a single path, the diffusion kernel-based method incorporates distances along multiple paths. 
Like the shortest paths method, the diffusion kernel-based method can also be applied to networks with or without edge weights. 
One interpretation of the method is the continuous time limit of a random walk across the network, resulting in highly-connected nodes being associated with smaller node pair distances. 

The negative graph Laplacian ($H$) is used to create a diffusion kernel for the network. $H$ is a square matrix of size $V \times V$ with entries:

\begin{equation}
H_{ij}^\text{unweighted} =   \begin{cases} \ 1 & \text{\small when}\  i \sim j \\ \ -d_i & \text{\small when}\  i=j \\ \ 0 & \text{\small when}\ i~/~j \end{cases} 
\end{equation}

\begin{equation}
H_{ij}^\text{weighted} = \  \begin{cases} \ w_{ij} & \text{\small when}\  i \sim j \\ \ - \sum_jw_{ij} & \text{\small when}\  i=j \\ \ 0 & \text{\small when}\ i~/~j, \end{cases} 
\end{equation} 

$H$ is specified for networks with and without edge weights. 
$i \sim j$ indicates that node $i$ and node $j$ are connected by an edge and $i~/~j$ indicates that they are not directly connected. 
$d_{i}$ is the number of edges associated with node $i$ (the degree of node $i$). 
$w_{ij}$ is the weight of the edge connecting nodes $i$ and $j$. 
The diffusion kernel can then be defined by calculating the matrix exponential ($D$): 

\begin{equation}
D =  \lim_{n \rightarrow \infty} \left( 1 + \frac{\beta H}{n} \right)^n = \exp ( \beta H ) 
\end{equation}

The fact that $H$ is diagonalizable ($H = U \Delta U^{-1}$) makes it easier to compute $D$. 
If $\Delta$ is a diagonal matrix with entries $(\delta_i)_{i=1, \ldots ,n}$, then $D = \exp ( \beta H) = U \exp (\beta \Delta) U^{-1}$. 
$\exp (\beta \Delta)$ is a diagonal matrix with entries $(\exp (\beta \delta_i))_{i=1, \ldots ,n}$ \cite{Kondor2002}.

\subsection*{Mean first-passage time-based distance measure} 
Mean first-passage time (MFPT) can also be used to compute the distances between pairs of nodes \cite{White2003}. 
The MFPT-based measure is similar to the diffusion kernel-based measure in that it can be compared to completing a random walk across the network. 
The MFPT of a walk from node $i$ to node $j$ ($m_{i,j}$)  represents the expected number of steps required to reach node $j$ for the first time:

\begin{equation}
\label{eq:mfpt}
m_{i,j} = \sum^{\infty}_{n=1} n~f^{(n)}_{i,j}
\end{equation} 

where $f^{(n)}_{i,j}$ is the probability that the random walk reaches node $j$ for the first time after $n$ steps. 
The MFPT between each node pair can be computed analytically using the equations:

\begin{equation}
\label{eq:mfpt_calc1}
\textbf{M} = (\textbf{I} - \textbf{Z} + \textbf{EZ}_{dg}) ~\textbf{D}
\end{equation}

\begin{equation}
\label{eq:mfpt_calc2}
\textbf{Z} = (\textbf{I} - \bf{e~\pi^{T}} - \textbf{A}) ^{-1}
\end{equation} 

where \textbf{I} is the identity matrix, \textbf{E} is a matrix with equal dimensions containing only 1s, \textbf{e} is a column vector containing only 1s, $\pi$ is a column vector of the stationary distributions of the Markov chain, \textbf{A} is the Markov chain transition matrix  and \textbf{D} is a diagonal matrix with elements:

\begin{equation}
\label{eq:mfpt_calc3}
d_{vv} = \frac{1}{\pi(v)}
\end{equation}

\subsection*{Costanzo et al. yeast GI networks} 
Costanzo et al. tested for genetic interactions (GI) between 5.4 million gene pairs in \emph{S. cerevisiae} using synthetic genetic array (SGA) analysis \cite{Costanzo2010}. 
Using this data, we created two interaction networks: the first from raw GI scores ($\epsilon$) and the second from correlations in interaction profile. 
The raw interaction network contains both positive ($\epsilon > 0.16$) and negative ($\epsilon < -0.12$) interactions (78,701 interactions between 4,326 genes). 
GI scores were converted into edge distances by calculating:

\begin{equation}
\label{eq:costanzo_raw}
d_e = -log_{10} \frac{ | \epsilon | } { | \epsilon | _{max}}
\end{equation}

The correlation network was created be computing, for each gene pair, Pearson's correlation coefficient for the respective rows of the complete GI matrix.
Pairs of genes were connected in the network if their interaction profile correlation coefficient exceeded a threshold.
Using a threshold of $PCC > 0.125$ ensured that the correlation network contained a similar number of interactions to the raw network (76,434 interactions between 4,326 genes).
Correlation coefficients ($c_e$) were converted into edge distances by calculating:

\begin{equation}
\label{eq:costanzo_correlation}
d_e = -log_{10}~c_e
\end{equation}


\subsection*{Bandyopadhyay et al. yeast GI networks} 
174,000 gene pairs were tested for interactions in MMS-treated and untreated \emph{S. cerevisiae} \cite{Bandyopadhyay2010}. 
Modified T-tests were used to compare the growth rate of the observed double mutant against the rate expected given that no interaction exists. 
We previously demonstrated that a strong correlation in GI profiles is a greater indicator of shared function than raw interactions. 
Therefore, we created a correlation network for each condition by computing Pearson's correlation coefficient for each gene pair. 
A threshold of $PCC > 0.3$ for the MMS-treated network and $PCC > 0.25942$ for the untreated network was applied to ensure that each network contained an equal number of interactions (3067 interactions between 419 genes). 
Correlation coefficients were converted into edge distances using Equation~\ref{eq:costanzo_correlation}.

\subsection*{Srivas et al. yeast GI networks} 
45,000 gene pairs were tested for interaction in \emph{S. cerevisiae} treated with high doses of UV radiation ($80~J/m^2$) and untreated \emph{S. cerevisiae}.
Modified T-tests were used to produce interaction scores ($S$) for each of the gene pairs.
Too few gene pairs were tested to build a GI correlation network and therefore networks of raw interactions were created. 
Pairs of genes were connected in the networks if $S >1.25$ or $S<-1.25$. 
The UV-treated network contains 5,799 interactions between 1,406 genes and the untreated network contains 6,270 interactions between 1,406 genes.
Interaction scores were converted into edge distances by calculating:
 
 \begin{equation}
\label{eq:srivas_raw}
d_e = -log_{10} \frac{ | S | } { | S | _{max}}
\end{equation}

\subsection*{IntAct physical and genetic interaction network}
\begin{figure}[!ht]
\begin{center}
\includegraphics[width=8cm, trim= 0mm 0mm 0mm 0mm]{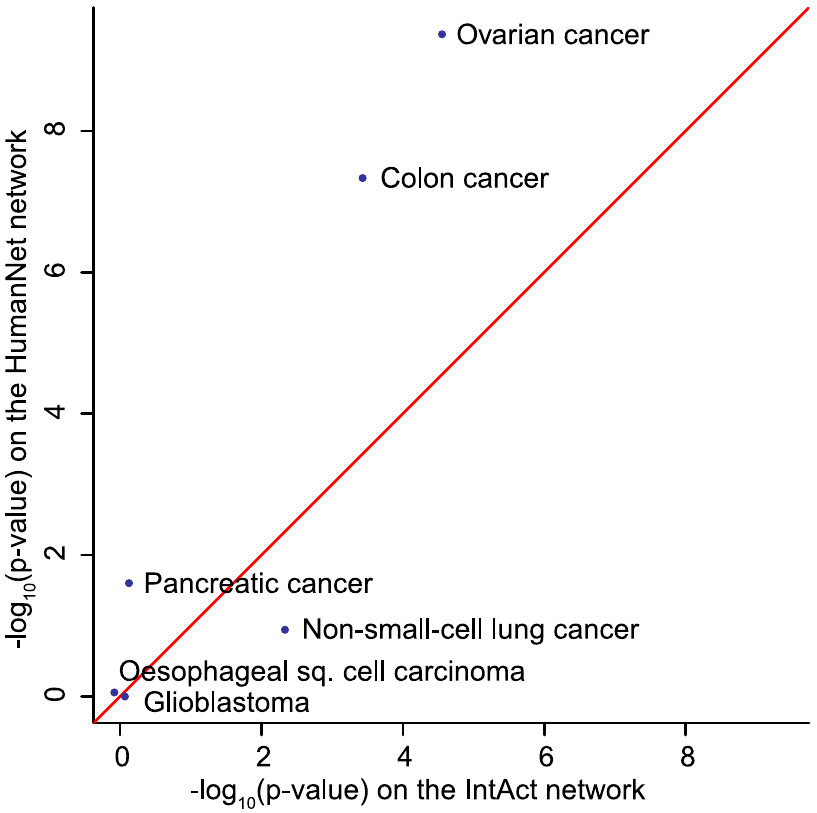}
\end{center}
\caption{
{\bf \Knet~identifies the most functionally informative network.}
Association of genes essential in the proliferation of cancer cell lines with a network of curated physical interactions (IntAct) and a functional network created using 21 data sources (HumanNet). 
Association was stronger between colon and ovarian cancer cell line RNAi hits and the functional network, indicating that the functional network provides more information about the mechanisms that drive cancer cell line maintenance.
}
\label{fig:rnai}
\end{figure}

IntAct is an open source database for molecular interaction data \cite{Orchard2014}.
\emph{H. sapien} data from the database was downloaded on 2013-05-02 to create the biological network used in Figure~\ref{fig:rnai}.
This network contains 6,856 genes and 21,291 interactions. 
No confidence scores were available for the interactions and therefore no edge distances are associated with the network.

\subsection*{HPRD physical interaction network}
The Human Protein Reference Database is a database of physical and functional interactions between genes and proteins \cite{KeshavaPrasad2009}. 
The HPRD network was downloaded from the R package DLBCL, version 1.3.7 \cite{DLBCL}. 
To allow for comparison of the \Knode~function to BioNet, only the largest cluster of interacting genes was used. 
The final HPRD network contains 7,756 interactions between 2,034 genes.

\subsection*{HumanNet functional interaction network}
HumanNet is a functional network that combines 21 sources of genomic and proteomic data from four species to build a human-specific biological network \cite{Lee2011}.
These sources of data include gene co-citation, gene co-expression, curated physical and genetic interactions, high-throughput physical and genetic interactions, co-occurrence of protein domains and bacterial orthologs from \emph{C. elegans}, \emph{D. melanogaster}, \emph{H. sapien} and \emph{S. cerevisiae}. 
Version 1 of the database was used to create the biological network used in Figure~\ref{fig:rnai}. 
Log likelihood scores were provided for each of the interactions. 
To reduce the density of the network, interactions with log likelihood scores less than 2 were removed from the network. 
This network contains 8,475 genes and 58,636 interactions. 
Log likelihood scores $LLS_e$ were converted into edge distances by calculating:  

\begin{equation}
\label{eq:humannet_distances}
d_e = -log_{10}\frac{LLS_e}{LLS_{e~max}}
\end{equation}

\subsection*{Cancer cell line RNAi hits}
RNAi technology can be used to identify genes essential to the survival of cancer cell lines. 
Cheung et al. performed genome-wide RNAi screens of 102 cell lines across 6 cancer types: oesophageal squamous cell carcinoma, glioblastoma (GBM), non-small-cell-lung cancer (NSCLC), pancreatic cancer, ovarian cancer and colon cancer \cite{Cheung2011}. 
11,194 genes were targeted. 
The weight of evidence approach was used to compute essentiality scores for each shRNA for each set of cancer cell lines \cite{Cheung2011}. 
GENE-E (\texttt{http://www.broadinstitute.org/cancer/software/GENE-E/index.html}) was used to collapse the shRNA-wise essentiality scores into gene-wise p-values. 
P-values are produced by permuting the shRNA scores 10,000 times in order to create artificial genes. 
The second best score of the shRNA within these artificial genes is then compared to the second best observed shRNA score. 
Gene-wise p-values $s_v$ were converted into node weights $p_v$ by calculating:

\begin{equation}
\label{eq:rnai_weights}
p_v = -log_{10}~s_v
\end{equation}

\subsection*{DLBCL gene expression and survival data}
Gene expression data for two subtypes of diffuse large B-cell lymphomas (DLBCL): germinal center B-like phenotype (GCB, 112 tumors) and activated B-like phenotype (ABC, 82 tumours), was obtained from the R package DLBCL, version 1.3.7 \cite{DLBCL}.
This package also contains data on patient survival. 
The data originally comes from a study of patient survival after chemotherapy \cite{Rosenwald2002a}. 
P-values for differential expression and risk association were produced using Cox regression. 
These p-values were combined using second-order statistics in order to produce gene-wise association scores which could be applied to the networks.
Gene-wise p-values were converted in to node weights using Equation~\ref{eq:rnai_weights}.

\subsection*{Gene Ontology database}
The Gene Ontology (GO) database consists of a hierarchical structure of gene annotations \cite{Ashburner2000}. 
Annotations from this database were used in Figure~\ref{fig:results1}. 
The GO database consists of 3 top-level ontologies: molecular functions, biological processes and cellular components, all of which were used in each figure. 
\emph{S. cerevisiae} GO term annotations were retrieved from the \emph{Saccharomyces} Genome Database (\texttt{www.yeastgenome.org}) using the R package org.Sc.sgd.db, version 2.10.1 \cite{org.Sc.sgd.db}.

\subsection*{Compactness score}
The compactness score $C$ is defined as the mean shortest path distance between pairs of nodes in a set $P$ on graph $G$ \cite{Glaab2010}. 

\begin{equation}
\label{eq:compactness}
C(P) = \frac{2  \sum_{i,j \in P; i < j} d^g(i,j)}{|P| * (|P| - 1)}
\end{equation} 

In order to measure the significance of the observed compactness score, we compared it to scores produced using sets of randomly permuted hits.
An empirical p-value for the observed compactness score is calculated using a Z-test.

\subsection*{Implementation}
The methodology described in this work has been assembled as an R package called SANTA, which is available for download at \texttt{http://bioconductor.org/packages/release/bioc/html/SANTA.html}. 
This package is distributed with the code (in the form of a vignette) and the data required to reproduce all of the results given in this paper. 
The vignette also contains the parameters used with the Barabasi-Albert model of preferential attachment to create the simulated networks.
The running time of SANTA depends on the size of the network and the number of permutations being run. 
Using 1000 permutations, SANTA requires 1GB of RAM and 25 seconds on a single Intel Xeon E5-2640 to measure the strength of association of a single gene set on the raw Costanzo et al. GI network (78,701 interactions between 4,326 genes). 
SANTA can use parallel computing (where available) to reduce running time.

\section*{Acknowledgments}
We acknowledge support by the University of Cambridge, Cancer Research UK, and Hutchison Whampoa Limited.

\bibliography{library}

\begin{thebibliography}{10}
\providecommand{\url}[1]{\texttt{#1}}
\providecommand{\urlprefix}{URL }
\expandafter\ifx\csname urlstyle\endcsname\relax
  \providecommand{\doi}[1]{doi:\discretionary{}{}{}#1}\else
  \providecommand{\doi}{doi:\discretionary{}{}{}\begingroup
  \urlstyle{rm}\Url}\fi
\providecommand{\bibAnnoteFile}[1]{%
  \IfFileExists{#1}{\begin{quotation}\noindent\textsc{Key:} #1\\
  \textsc{Annotation:}\ \input{#1}\end{quotation}}{}}
\providecommand{\bibAnnote}[2]{%
  \begin{quotation}\noindent\textsc{Key:} #1\\
  \textsc{Annotation:}\ #2\end{quotation}}
\providecommand{\eprint}[2][]{\url{#2}}

\bibitem{Ashburner2000}
Ashburner M, Ball CA, Blake JA, Botstein D, Heather B, et~al. (2000) {Gene
  Ontology: tool for the unification of biology}.
\newblock Nature Genetics 25: 25--29.
\bibAnnoteFile{Ashburner2000}

\bibitem{Subramanian2005}
Subramanian A, Tamayo P, Mootha VK, Mukherjee S, Ebert BL, et~al. (2005) {Gene
  Set Enrichment Analysis: a Knowledge-Based Approach for Interpreting
  Genome-Wide Expression Profiles}.
\newblock PNAS 102: 15545--50.
\bibAnnoteFile{Subramanian2005}

\bibitem{Beissbarth2004}
Beissbarth T, Speed TP (2004) {GOstat: find statistically overrepresented Gene
  Ontologies within a group of genes}.
\newblock Bioinformatics 20: 1464--5.
\bibAnnoteFile{Beissbarth2004}

\bibitem{Khatri2012}
Khatri P, Sirota M, Butte AJ (2012) {Ten years of pathway analysis: current
  approaches and outstanding challenges}.
\newblock PLOS Computational Biology 8: e1002375.
\bibAnnoteFile{Khatri2012}

\bibitem{Vidal2011}
Vidal M, Cusick ME, Barab\'{a}si AL (2011) {Interactome networks and human
  disease}.
\newblock Cell 144: 986--98.
\bibAnnoteFile{Vidal2011}

\bibitem{Rozenblatt-Rosen2012}
Rozenblatt-Rosen O, Deo RC, Padi M, Adelmant G, Calderwood Ma, et~al. (2012)
  {Interpreting cancer genomes using systematic host network perturbations by
  tumour virus proteins}.
\newblock Nature 487: 491--5.
\bibAnnoteFile{Rozenblatt-Rosen2012}

\bibitem{Costanzo2010}
Costanzo M, Baryshnikova A, Bellay J, Kim Y, Spear ED, et~al. (2010) {The
  Genetic Landscape of a Cell}.
\newblock Science 327: 425--31.
\bibAnnoteFile{Costanzo2010}

\bibitem{Ryan2012}
Ryan CJ, Roguev A, Patrick K, Xu J, Jahari H, et~al. (2012) {Hierarchical
  modularity and the evolution of genetic interactomes across species}.
\newblock Molecular Cell 46: 691--704.
\bibAnnoteFile{Ryan2012}

\bibitem{Laufer2013}
Laufer C, Fischer B, Billmann M, Huber W, Boutros M (2013) {Mapping genetic
  interactions in human cancer cells with RNAi and multiparametric
  phenotyping}.
\newblock Nature Methods 10: 427--31.
\bibAnnoteFile{Laufer2013}

\bibitem{Bandyopadhyay2010}
Bandyopadhyay S, Mehta M, Kuo D, Sung Mk, Chuang R, et~al. (2010) {Rewiring of
  genetic networks in response to DNA damage}.
\newblock Science 330: 1385--1390.
\bibAnnoteFile{Bandyopadhyay2010}

\bibitem{Guenole2013}
Gu\'{e}nol\'{e} A, Srivas R, Vreeken K, Wang ZZ, Wang S, et~al. (2013)
  {Dissection of DNA damage responses using multiconditional genetic
  interaction maps}.
\newblock Molecular Cell 49: 346--58.
\bibAnnoteFile{Guenole2013}

\bibitem{Kiel2013}
Kiel C, Ebhardt H, Burnier J, Portugal C, Sabido E, et~al. (2013)
  {Quantification of ErbB Network Proteins in Three Cell Types Using
  Complementary Approaches Identifies Cell-General and Cell-Type-Specific
  Signaling Proteins}.
\newblock Journal of Proteome Research 13: 300--13.
\bibAnnoteFile{Kiel2013}

\bibitem{Wang2006}
Wang J, Rao S, Chu J, Shen X, Levasseur DN, et~al. (2006) {A protein
  interaction network for pluripotency of embryonic stem cells.}
\newblock Nature 444: 364--8.
\bibAnnoteFile{Wang2006}

\bibitem{Pujana2007}
Pujana MA, Han JDJ, Starita LM, Stevens KN, Tewari M, et~al. (2007) {Network
  modeling links breast cancer susceptibility and centrosome dysfunction.}
\newblock Nature Genetics 39: 1338--49.
\bibAnnoteFile{Pujana2007}

\bibitem{Sakai2011}
Sakai Y, Shaw Ca, Dawson BC, Dugas DV, Al-Mohtaseb Z, et~al. (2011) {Protein
  interactome reveals converging molecular pathways among autism disorders.}
\newblock Science Translational Medicine 3: 86ra49.
\bibAnnoteFile{Sakai2011}

\bibitem{Ripley1981}
Ripley BD (1981) {Spatial Statistics}.
\newblock Hoboken, NJ: John Wiley \& Sons, Inc.
\bibAnnoteFile{Ripley1981}

\bibitem{Rahnenfeuhrer2004}
Rahnenf\"{u}hrer J, Domingues FS, Maydt J, Lengauer T (2004) {Calculating the
  statistical significance of changes in pathway activity from gene expression
  data}.
\newblock Statistical Applications in Genetics and Molecular Biology 3: Article
  16.
\bibAnnoteFile{Rahnenfeuhrer2004}

\bibitem{Draghici2007}
Draghici S, Khatri P, Tarca AL, Amin K, Done A, et~al. (2007) {A systems
  biology approach for pathway level analysis}.
\newblock Genome Research 17: 1537--45.
\bibAnnoteFile{Draghici2007}

\bibitem{Shojaie2010}
Shojaie A, Michailidis G (2010) {Network Enrichment Analysis in Complex
  Experiments}.
\newblock Statistical Applications in Genetics and Molecular Biology 9: Article
  22.
\bibAnnoteFile{Shojaie2010}

\bibitem{Glaab2010}
Glaab E, Baudot A, Krasnogor N, Valencia A (2010) {Extending pathways and
  processes using molecular interaction networks to analyse cancer genome
  data.}
\newblock BMC Bioinformatics 11: 597.
\bibAnnoteFile{Glaab2010}

\bibitem{Glaab2012}
Glaab E, Baudot A, Krasnogor N, Schneider R, Valencia A (2012) {EnrichNet:
  network-based gene set enrichment analysis}.
\newblock Bioinformatics 28: i451--7.
\bibAnnoteFile{Glaab2012}

\bibitem{Ideker2002}
Ideker T, Ozier O, Schwikowski B, Siegel A (2002) {Discovering regulatory and
  signalling circuits in molecular interaction networks}.
\newblock Bioinformatics 18: 233--240.
\bibAnnoteFile{Ideker2002}

\bibitem{Sanguinetti2008}
Sanguinetti G, Noirel J, Wright PC (2008) {MMG: a probabilistic tool to
  identify submodules of netabolic pathways}.
\newblock Bioinformatics 24: 1078--1084.
\bibAnnoteFile{Sanguinetti2008}

\bibitem{Beisser2010}
Beisser D, Klau GW, Dandekar T, M\"{u}ller T, Dittrich MT (2010) {BioNet: an
  R-Package for the functional analysis of biological networks}.
\newblock Bioinformatics 26: 1129--30.
\bibAnnoteFile{Beisser2010}

\bibitem{Jia2011}
Jia P, Zheng S, Long J, Zheng W, Zhao Z (2011) {dmGWAS: dense module searching
  for genome-wide association studies in protein-protein interaction networks}.
\newblock Bioinformatics 27: 95--102.
\bibAnnoteFile{Jia2011}

\bibitem{Gaetan2010}
Gaetan C, Guyon X (2010) {Spatial Statistics and Modeling}.
\newblock New York, NY: Springer.
\bibAnnoteFile{Gaetan2010}

\bibitem{Suratanee2010}
Suratanee A, Rebhan I, Matula P, Kumar A, Kaderali L, et~al. (2010) {Detecting
  host factors involved in virus infection by observing the clustering of
  infected cells in siRNA screening images}.
\newblock Bioinformatics 26: i653--8.
\bibAnnoteFile{Suratanee2010}

\bibitem{Kondor2002}
Kondor RI, Lafferty J (2002) {Diffusion kernels on graphs and other discrete
  structures}.
\newblock Proceedings of the Nineteenth International Conference on Machine
  Learning : 315--22.
\bibAnnoteFile{Kondor2002}

\bibitem{White2003}
White S, Smyth P (2003) {Algorithms for estimating relative importance in
  networks}.
\newblock Proceedings of the Ninth ACM SIGKDD International Conference on
  Knowledge Discovery and Data Mining : 266--75.
\bibAnnoteFile{White2003}

\bibitem{Cormen2009}
Cormen TH, Leiserson CE, Rivest RL, Stein C (2009) {Introduction to
  Algorithms}.
\newblock Cambridge, MA: MIT Press.
\bibAnnoteFile{Cormen2009}

\bibitem{Csardi2006}
Csardi G, Napusz T (2006) {The igraph software package for complex network
  research}.
\newblock Inter Journal Complex Systems : 1695.
\bibAnnoteFile{Csardi2006}

\bibitem{Barabasi1999}
Barab\'{a}si A, Albert R (1999) {Emergence of Scaling in Random Networks}.
\newblock Science 286: 509--512.
\bibAnnoteFile{Barabasi1999}

\bibitem{Rosenwald2002}
Rosenwald A, Wright G, Chan WC, Connors JM, Campo E, et~al. (2002) {The use of
  molecular profiling to predict survival after chemotherapy for diffuse
  large-B-cell lymphoma.}
\newblock The New England Journal of Medicine 346: 1937--47.
\bibAnnoteFile{Rosenwald2002}

\bibitem{KeshavaPrasad2009}
{Keshava Prasad} TS, Goel R, Kandasamy K, Keerthikumar S, Kumar S, et~al.
  (2009) {Human Protein Reference Database-2009 update}.
\newblock Nucleic Acids Research 37: D767--72.
\bibAnnoteFile{KeshavaPrasad2009}

\bibitem{Huang2009a}
Huang DW, Sherman BT, Lempicki Ra (2009) {Systematic and integrative analysis
  of large gene lists using DAVID bioinformatics resources.}
\newblock Nature Protocols 4: 44--57.
\bibAnnoteFile{Huang2009a}

\bibitem{Frost2012}
Frost A, Elgort MG, Brandman O, Ives C, Collins SR, et~al. (2012) {Functional
  repurposing revealed by comparing S. pombe and S. cerevisiae genetic
  interactions}.
\newblock Cell 149: 1339--52.
\bibAnnoteFile{Frost2012}

\bibitem{StOnge2007}
{St Onge} R, Mani R, Oh J, Proctor M, Fung E, et~al. (2007) {Systematic pathway
  analysis using high-resolution fitness profiling of combinatorial gene
  deletions}.
\newblock Nature Genetics 39: 199--206.
\bibAnnoteFile{StOnge2007}

\bibitem{Srivas2013a}
Srivas R, Costelloe T, Carvunis AR, Sarkar S, Malta E, et~al. (2013) {A
  UV-induced genetic network links the RSC complex to nucleotide excision
  repair and shows dose-dependent rewiring.}
\newblock Cell Reports 5: 1714--24.
\bibAnnoteFile{Srivas2013a}

\bibitem{Sertic2012}
Sertic S, Pizzi S, Lazzaro F, Plevani P, Muzi-falconi M (2012) {NER and DDR.
  Classical music with new instruments}.
\newblock Cell Cycle 11: 668--674.
\bibAnnoteFile{Sertic2012}

\bibitem{Tham2002}
Tham W, Zakian V (2002) {Transcriptional silencing at Saccharomyces telomeres:
  implications for other organisms.}
\newblock Oncogene 21: 512--521.
\bibAnnoteFile{Tham2002}

\bibitem{Martin1999}
Martin SG, Laroche T, Suka N, Grunstein M, Gasser SM (1999) {Relocalization of
  telomeric Ku and SIR proteins in response to DNA strand breaks in yeast.}
\newblock Cell 97: 621--33.
\bibAnnoteFile{Martin1999}

\bibitem{Cheung2011}
Cheung HW, Cowley GS, Weir Ba, Boehm JS, Rusin S, et~al. (2011) {Systematic
  investigation of genetic vulnerabilities across cancer cell lines reveals
  lineage-specific dependencies in ovarian cancer}.
\newblock PNAS 108: 12372--7.
\bibAnnoteFile{Cheung2011}

\bibitem{Orchard2014}
Orchard S, Ammari M, Aranda B, Breuza L, Briganti L, et~al. (2014) {The MIntAct
  project--IntAct as a common curation platform for 11 molecular interaction
  databases}.
\newblock Nucleic Acids Research 42: D358--63.
\bibAnnoteFile{Orchard2014}

\bibitem{Lee2011}
Lee I, Blom UM, Wang PI, Shim JE, Marcotte EM (2011) {Prioritizing candidate
  disease genes by network-based boosting of genome-wide association data}.
\newblock Genome Research 21: 1109--21.
\bibAnnoteFile{Lee2011}

\bibitem{Tong2004}
Tong AHY, Lesage G, Bader GD, Ding H, Xu H, et~al. (2004) {Global mapping of
  the yeast genetic interaction network}.
\newblock Science 303: 808--13.
\bibAnnoteFile{Tong2004}

\bibitem{Barabasi2011}
Barab\'{a}si AL, Gulbahce N, Loscalzo J (2011) {Network medicine: a
  network-based approach to human disease}.
\newblock Nature Reviews Genetics 12: 56--68.
\bibAnnoteFile{Barabasi2011}

\bibitem{Creixell2012}
Creixell P, Schoof EM, Erler JT, Linding R (2012) {Navigating cancer network
  attractors for tumor-specific therapy}.
\newblock Nature Biotechnology 30: 842--8.
\bibAnnoteFile{Creixell2012}

\bibitem{Erler2012}
Erler JT, Linding R (2012) {Network medicine strikes a blow against breast
  cancer}.
\newblock Cell 149: 731--3.
\bibAnnoteFile{Erler2012}

\bibitem{Hofree2013}
Hofree M, Shen JP, Carter H, Gross A, Ideker T (2013) {Network-based
  stratification of tumor mutations}.
\newblock Nature Methods 10: 1108--15.
\bibAnnoteFile{Hofree2013}

\bibitem{DLBCL}
Dittrich M, Beisser D (2010).
\newblock {DLBCL: Diffuse large B-cell lymphoma expression data}.
\newblock \urlprefix\url{http://bionet.bioapps.biozentrum.uni-wuerzburg.de/}.
\bibAnnoteFile{DLBCL}

\bibitem{Rosenwald2002a}
Rosenwald A, Wright G, Chan WC, Connors JM, Campo E, et~al. (2002) {The use of
  molecular profiling to predict survival after chemotherapy for diffuse
  large-B-cell lymphoma.}
\newblock The New England journal of medicine 346: 1937--47.
\bibAnnoteFile{Rosenwald2002a}

\bibitem{org.Sc.sgd.db}
Carlson M.
\newblock {org.Sc.sgd.db: Genome wide annotation for Yeast}.
\newblock
  \urlprefix\url{http://www.bioconductor.org/packages/release/data/annotation/html/org.Sc.sgd.db.html}.
\bibAnnoteFile{org.Sc.sgd.db}

\end{thebibliography}

\clearpage
\section*{Figure Legends}

\begin{suppfig}[!ht]
\begin{center}
\includegraphics[width=8cm, trim= 0mm 0mm 0mm 0mm]{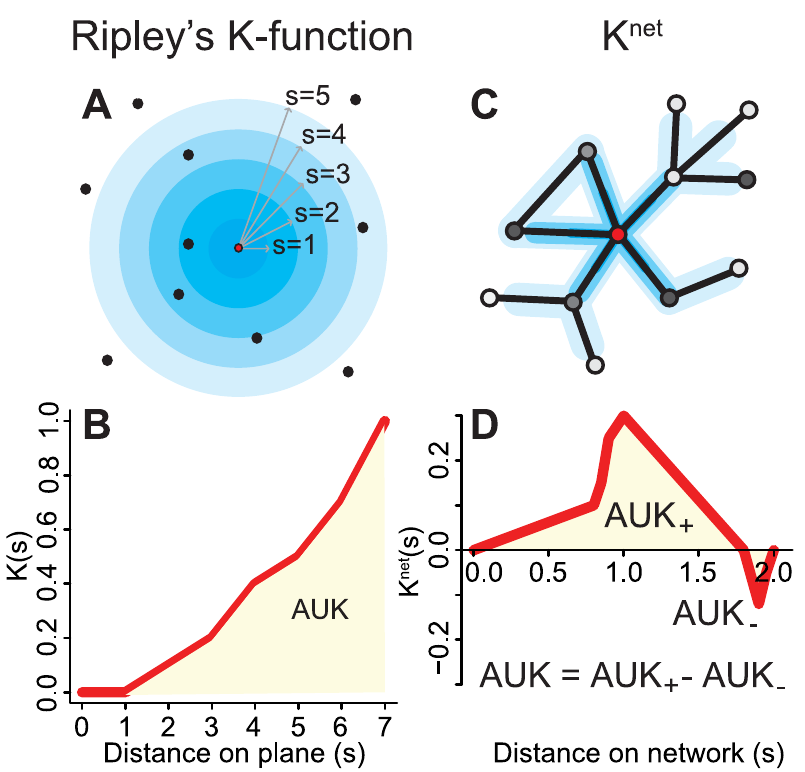}
\end{center}
\caption{
{\bf \bf Comparison of Ripley's K-function and \Knet.}
(A) Ripley's K-function ($K(s)$) counts how many points on a plane are captured within circles of increasing radius ($s$) around each point. 
Here, circles are drawn from only a single point (red circle). 
(B) The graph of $K(s)$ for the distribution of points in (A). 
If the clustering of points were greater, then the K(s) function would increase faster and the area under the curve (AUK) would be greater. 
(C) The \Knet-function computes the absolute deviation of the sum of the weight of nodes within a certain distance of each node from the Null model. 
The distance from a single node (red circle) is shown. 
The darker the colour of the node, the greater its weight. 
(D) The graph of the \Knet-function for the network and node weights in (C). 
The greater the clustering of the node weights on the network, the greater the AUK.}
\label{suppfig:ripley_vs_knet}
\end{suppfig}

\begin{suppfig}[!ht]
\begin{center}
\includegraphics[width=8cm, trim= 0mm 0mm 0mm 0mm]{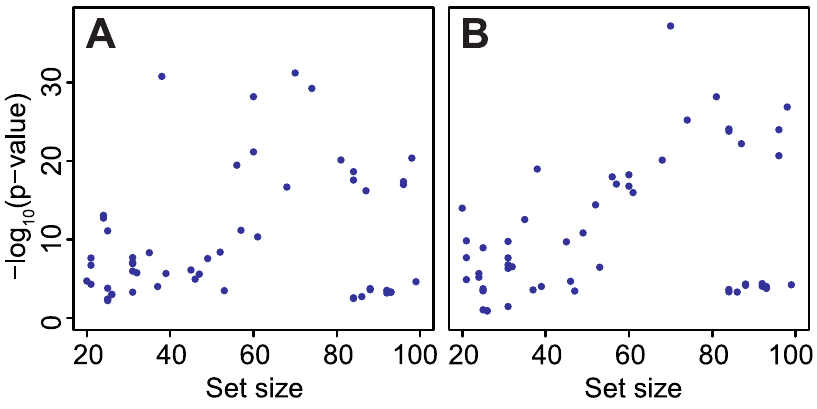}
\end{center}
\caption{
{\bf Correlation between set size and \Knet{} p-value.}
Plot of the significance of the clustering of sets of network genes associated with a GO term against the set size on GI networks mapped in (A) untreated yeast and (B) yeast treated with the DNA-damaged agent MMS. 
Only those GO terms that associate with either or both networks with a strength of $p  < 0.001$ are shown. 
Many GO terms share a large number of genes due to their ontological relationship. 
When those GO terms that are ancestors of other GO terms tested are removed, Pearson's correlation coefficient equals 0.004 for the treated network and -0.040 for the untreated network, demonstrating that there is little correlation between set size and \Knet{} p-value.
}
\label{suppfig:clustering_vs_size}
\end{suppfig}

\begin{suppfig}[!ht]
\includegraphics[width=14cm, trim= 0mm 0mm 0mm 0mm]{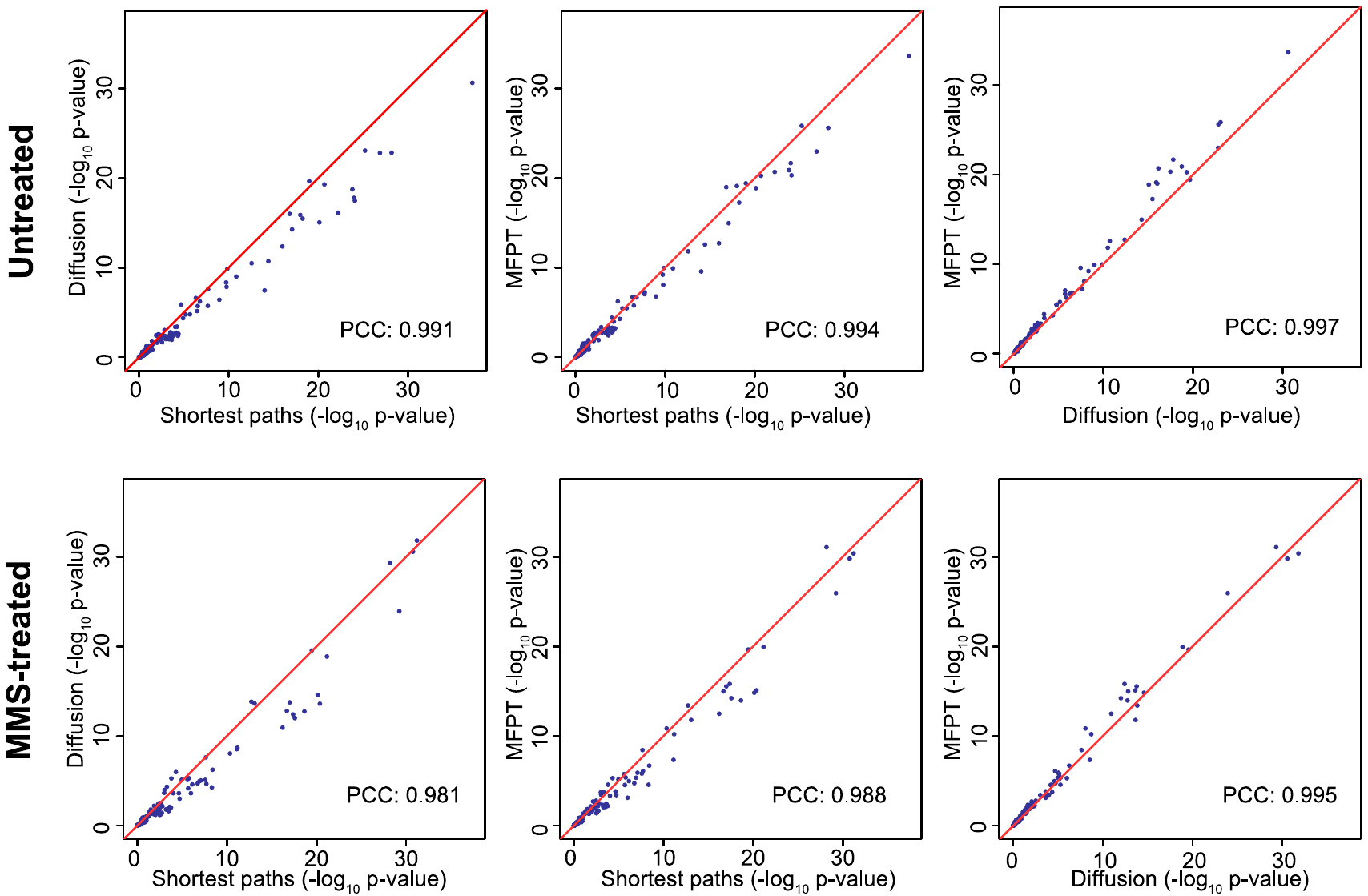}
\begin{center}
\end{center}
\caption{
{\bf Correlation in network-gene set association strength between distance methods.} 
Pair-wise comparison of the association strengths of GO terms across the three distance methods. The networks tested were the MMS-treated (Top) and untreated (Bottom) \emph{S. cerevisiae} GI networks created using data from Bandyopadhyay et al. Association strength correlation across networks is very high ($PCC >0.98$), demonstrating that the results produced by SANTA are generally robust across distance methods.
}
\label{suppfig:distance_robustness}
\end{suppfig}

\clearpage
\section*{Tables}

\begin{supptable}[!ht]
\caption{
{\bf GO terms differentially associated with a network of raw GIs and GI profile correlations.}
\Knet{} was used to test the strength of association between sets of genes associated with various GO terms and the two network types. 
This table contains the GO terms that associated most strongly ($p < 1 \times 10^{-8}$) with one or both of the networks. 
GO terms are ranked by their differential association strength ($D$), with the terms associated more strongly with the network of GI profile correlations positioned towards the top and the terms associated more strongly with the network of raw GIs positioned towards the bottom. 
A greater number of GO term genes associated more strongly with the network of GI profile correlations. 
}
\label{supptable:raw_correlation}
 \end{supptable}

\begin{supptable}[!ht]
\caption{
{\bf GO terms differentially associated with the untreated and MMS-treated GI networks.}
\Knet{} was used to test the strength of association between sets of genes associated with various GO terms and the two network types. 
The table contains the GO terms that associated most strongly ($p < 0.001$) with one or both of the networks. 
GO terms are ranked by their differential association strength ($D$), with the terms associated more strongly with the treated network positioned towards the top and the terms associated more strongly with the untreated network positioned towards the bottom. 
}
\label{supptable:dna_damage}
 \end{supptable}

\begin{supptable}[!ht]
\caption{
{\bf GO terms differentially associated with the untreated and UV-treated GI networks.}
\Knet{} was used to test the strength of association between sets of genes associated with various GO terms and the two network types. 
The table contains the GO terms that associated most strongly ($p < 0.001$) with one or both of the networks. 
GO terms are ranked by their differential association strength ($D$), with the terms associated more strongly with the treated network positioned towards the top and the terms associated more strongly with the untreated network positions towards the bottom. 
}
\label{supptable:uv}
 \end{supptable}

\end{document}